\documentclass[twocolumn,prd,showpacs]
              {revtex4}
\usepackage{graphicx,color}

\begin{document}

\title{Dust-shell Universe in the modified gravity scenario}

\author{Michael~Maziashvili}
\email{maziashvili@ictsu.tsu.edu.ge} \affiliation{Department of
Theoretical Physics, Tbilisi State University, \\ 3 Chavchavadze
Ave., Tbilisi 0128, Georgia}

\begin{abstract}
The dynamics of the dust-shell model of universe is exactly solved
for the modified Schwarzschild solution. This solution is used to
derive the cosmology corresponding to the modified gravity.

\end{abstract}


\maketitle

\section{Introduction}

The CMBR gives a strong observational basis for the assumption of
homogeneity and isotropy of our universe. Indeed, the deviation of
universe from a homogenous and isotropic space is $\sim 10^{-5}$
at the stage of decoupling. Thus our universe is well approximated
by the Friedmann-Lema\^{i}tre model before this stage. On the
other hand the present universe is highly inhomogeneous on small
scales. The model consisting of the spherical shells of dust with
the common center, where the rest observer is situated, was
considered in \cite{SNH1,SNH2} to study how the small scale
inhomogeneities affect the global dynamics when averaged on larger
scales. The gravitational field in the region between the shells
is described due to Birkhoff's theorem by the standard
Schwarzschild geometry and Israel's \cite{Is} matching conditions
are used to derive the equations of motion of the dust shell. The
dynamics of this model is exactly integrable that allows one to
make the analytical evaluations. On the other hand in the last two
decades Milgrom's modified Newtonian dynamics (MOND) paradigm
\cite{Mi1} has gained recognition as a successful scheme for
unifying much of extragalactic dynamics phenomenology without
invoking dark matter \cite{MS}. By presuming the Birkhoff's law
the relation between modified gravity and corresponding cosmology
was established in \cite{LSS}. In this framework the viability of
MOND cosmology was considered in \cite{LS}. In this brief note we
consider the dust shell universe under assumption of modified
Schwarzschild solution. Throughout this paper we put $G=c=1$.

Let us begin by considering of a single dust shell under
assumption of modified Schwarzschild solution given by
\begin{equation}ds^2=f^{-1}dr^2+r^2(d\theta^2+\sin^2\theta
d\varphi^2)-fdt^2~,\end{equation} where $f=1-2g(m/r)$ with some
function $g$ differing from the standard Schwarzschild solution
$g(x)=x$.  Using Israel's matching conditions \cite{Is} the
equation of motion of a spherical shell of dust takes the form

\begin{eqnarray}\label{eqmot}&\ddot{R}&\left\{\sqrt{1+\dot{R}^2}+\sqrt{1+\dot{R}^2-2g\left(\frac{m}{R}\right)}\,\right\}=\nonumber\\
&-&g'\left(\frac{m}{R}\right)\frac{m\sqrt{1+\dot{R}^2}}{R^2}~,\end{eqnarray}
where the prime stands for the derivative with respect to the
whole argument and over-dot denotes derivative with respect to the
proper time $\tau$. The relation between $t$ and $\tau$ is
\begin{equation}\frac{dt}{d\tau}=[1-2g(m/R)]^{-1}\left[1-2g\left(\frac{m}{R}\right)+\left(\frac{dR}{d\tau}\right)^2\right]~.\end{equation}
The first integral of eq.(\ref{eqmot}) is given by
\begin{equation}\label{fint}(1+\dot{R}^2)^{1/2}=a+\frac{g(m/R)}{2a}~,\end{equation}
where $a$ is the constant of integration characterizing the
binding energy of the shell. From eq.(\ref{fint}) one sees that in
the case of standard Schwarzschild solution expanding shell with
positive binding energy corresponding to $a<1$ falls back after
reaching a finite maximal radius. In the case of MOND the gravity
at large distances behaves as \cite{LS}
\[g\left(\frac{m}{r}\right)\sim \ln \left(\frac{2m}{r^3H_0^2}\right)+const.~,\] or if accommodated to
the cosmological data \cite{LS}
\[g\left(\frac{m}{r}\right)\sim (rH_0)^2~,\]
where $H_0$ denotes the present value of the Hubble constant. So,
in the case of MOND the dust shell can tunnel through the
potential barrier to the regime of unbounded expansion.

Let us now consider the dust shell universe which represents the
spherically symmetric shells of dust with a common center where
the rest observer is situated. Denoting the mass and radius of the
$i$th shell by $m_i$ and $R_i$ the equation of motion of the $i$th
shell takes the form

\begin{eqnarray}\label{eqmottwo}\ddot{R_i}\left\{\left[1+\dot{R_i}^2-2g\left(\frac{m_{i-1}}{R_i}\right)\right]^{1/2}
\right.&+& \nonumber\\
\left.\left[1+\dot{R_i}^2-2g\left(\frac{m_{i}}{R_i}\right)\right]^{1/2}\,\right\}&=&\nonumber\\
-\frac{m_{i-1}g'(m_{i-1}/R_i)}{R_i^2}\left[1+\dot{R_i}^2-2g\left(\frac{m_{i}}{R_i}\right)\right]^{1/2}&&\nonumber\\
-\frac{m_{i}g'(m_{i}/R_i)}{R_i^2}\left[1+\dot{R_i}^2-2g\left(\frac{m_{i-1}}{R_i}\right)\right]^{1/2}~.&&\end{eqnarray}

The first integral of this equation can be written as

\begin{eqnarray}\label{sol}1+\dot{R_i}^2&=&a_i^2+g\left(\frac{m_{i-1}}{R_i}\right)+g\left(\frac{m_{i}}{R_i}\right)\nonumber
\\&+&\frac{1}{4a_i^2}\left[g\left(\frac{m_{i}}{R_i}\right)-g\left(\frac{m_{i-1}}{R_i}\right)\right]^2~.
\end{eqnarray}
Recalling that in the standard case the matter density $\rho_i$ is
defined as \cite{SNH1, SNH2}
\[\frac{m_{i-1}+m_{i}}{2}=\frac{4}{3}\pi\rho_ix_i^3~,\] where $x_i$ denotes the
initial radius of the $i$th shell, we find it natural to define
the matter density in the case of modified gravity as follows
\[\frac{1}{2}\left[g\left(\frac{m_{i-1}}{R}\right)+g\left(\frac{m_{i}}{R}\right)\right]=g\left(\frac{4}{3}\pi\rho_i\frac{x_i^3}{R}\right)~.\]
If one sets the parameters containing in eq.(\ref{sol}) as follows
\[\rho_i=\rho~,~~~a_i^2-1=-kx_i^2\] the expansion law of the dust
shell can be written as
\begin{eqnarray}\label{modexp}\left(\frac{\dot{R_i}}{R_i}\right)^2&=&-k\left(\frac{x_i}{R_i}\right)^2+\frac{2}{R_i^2}g\left(\frac{4}{3}\pi\rho\frac{x_i^3}{R_i}\right)\nonumber\\
&+&\frac{1}{4a_i^2R_i^2}\left[g\left(\frac{m_{i}}{R_i}\right)-g\left(\frac{m_{i-1}}{R_i}\right)\right]^2~.\end{eqnarray}
The last radiation like term in eq.(\ref{modexp}) can be neglected
in the large $N$ limit \cite{SNH1, SNH2}. Namely, defining
\[x_i=i\Delta x~,~~~N\Delta x=\sqrt{\frac{3}{8\pi\rho}}~,\]
where $N$ denotes the number of shells one finds

\begin{eqnarray*}m_{2n}-m_{2n-1}&=&\frac{n^2}{N^3}\sqrt{\frac{108}{8\pi\rho}}~,
\\m_{2n+1}-m_{2n}&=&\frac{6n^2+6n+1}{N^3}\sqrt{\frac{3}{8\pi\rho}}~.\end{eqnarray*}
So that for large enough values of $N$ the difference
$g\left(m_i/R_i\right)-g\left(m_{i-1}/R_i\right)$ becomes quite
small that makes it possible to omit this term in the large $N$
limit. Neglecting the last radiation like term and introducing a
new function $\lambda$ \cite{LS}

\[2g\left(\frac{2m}{r}\right)\equiv
r^2H_0^2\lambda\left(\frac{2m}{r^3H_0^2}\right)~,\]

from eq.(\ref{sol}) one finds
\begin{equation}\left(\frac{\dot{R}_i}{R_i}\right)^2=-k\left(\frac{x_i}{R_i}\right)^2+H_0^2\lambda\left(\frac{8\pi\rho}{3H_0^2}\,\frac{x_i^3}{R_i^3}\right)~,\end{equation}
which after redefinition $x_i/R_i\rightarrow \mbox{\emph{scale
factor}}$ determines the cosmology with an arbitrary curvature
term corresponding to the modified gravity \cite{LSS, LS}.

To summarize we presented exact solution of dust shell universe
dynamics in the case of modified gravity and by means of this
solution derived the equation governing the cosmological
evolution. Using the present solution one can continue in the
spirit of papers \cite{SNH1,SNH2} to study an impact of
small-scale inhomogeneities on the modified cosmology. However,
one has to keep in mind that this model is highly idealized. In
general local inhomogeneities strongly affect the light
propagation, giving rise to dispersions in the observed
distance-redshift relation.

\begin{acknowledgments}
The author is indebted to M.~Gogberashvili for useful
conversations. The work was supported by the grant REL.RIG.9807.
\end{acknowledgments}


\end{document}